\RequirePackage{ifpdf}
\documentclass[a4paper,11pt,hyper]{JHEP3}

\usepackage{amsfonts,latexsym,graphicx,epsfig,amssymb,amsmath,mathrsfs}

\newcommand{\bm}[1]{\hbox{\boldmath{$#1$}}}
\newcommand{\sbm}[1]{\hbox{\boldmath{\scriptsize$#1$}}}
\newcommand{\dd}{{\rm d}}

\newcommand{\WQ}{W_{{\rm QFT}_+}}

\newcommand{\aph}{\bm{\phi}}

\title{Multi-field inflation from holography}
\author{Jaume Garriga$^a$, Kostas Skenderis$^b$, Yuko Urakawa$^{a,c}$
\\
a. Departament de F{\'\i}sica Fonamental i Institut de Ci{\`e}ncies del Cosmos,
Universitat de Barcelona,
Mart{\'\i}\ i Franqu{\`e}s 1, 08028 Barcelona, Spain\\
b. STAG Research Centre and Mathematical Sciences, University of Southampton, Southampton SO17 1BJ, UK \\
c. Department of Physics and Astrophysics, Nagoya University, Chikusa,
Nagoya 464-8602, Japan}

\abstract{We initiate the study of multi-field inflation using holography.
Bulk light scalar fields correspond to nearly marginal operators in the boundary theory and 
the dual quantum field theory is a deformation of a CFT by such operators.
We compute the power spectra of adiabatic and entropy perturbations in a simple model and find that 
the adiabatic curvature perturbation is not conserved in the presence of entropy perturbations 
but becomes conserved when the entropy perturbations are set to zero or the model is effectively a single scalar model, in agreement with expectations from cosmological perturbation theory.}

\preprint{}

\maketitle

\begin{document}

\section{Introduction}
The physics of the very Early Universe provides a window to Planck scale
physics where quantum gravity effects are expected to play an important
role. Moreover, recent cosmological observations provide us with a unique opportunity to test theoretical
ideas about the physics of this period 
and strongly suggest that the Universe underwent a period of rapid accelerated expansion at early
times, a period of cosmic inflation. The underlying theoretical
description of inflation is based on perturbative quantization around an
accelerating Friedmann- Robertson-Walker (FRW) spacetime. This
description however is in general sensitive to the UV completion of the theory and requires an embedding
of inflation in a consistent quantum theory of gravity. Furthermore, the
degrees of freedom of the Early Universe may have been strongly coupled
requiring an altogether new framework for the description of their
dynamics.

Holography offers a framework that could accommodate both conventional
inflation and also new models describing a Universe that is
non-geometric and strongly interacting at early times. Recall that in
gauge/gravity duality~\cite{Maldacena1997, GKP, Witten1998} a strongly
coupled phase of bulk gravity is dual to a weakly coupled quantum field
theory (QFT) and vice versa. Thus conventional inflation could  have a
dual description in terms of a strongly coupled QFT while weakly coupled
QFT could describe  the dynamics of a non-geometric Universe.

The application of gauge/gravity correspondence to cosmology was initiated in
Refs.~\cite{Witten, Strominger} (see also Ref.~\cite{Hull:1998vg}) where a correspondence between
de Sitter (dS) and conformal field theory (CFT) was discussed. The
conjecture was further sharpened in Ref.~\cite{Maldacena02} where it was
argued that the CFT partition function computes the late time behavior
of the dS wavefunction upon a certain analytic continuation. Early
related work on dS and quasi-dS  (including discussions of cosmological
perturbations) can be found in
Refs.~\cite{Bousso:2001mw,Strominger2,Larsen:2002et, LM03, vdS, LM04,
Seery:2006tq} while more recent works include 
Refs.~\cite{Harlow:2011ke, Maldacena:2011nz,Maldacena:2011mk,Anninos:2011ui, Hertog:2011ky, Hartle:2012qb,Anninos:2012ft,
Hartle:2012tv, Shiu, Mata:2012bx, JYsingle, Hartle:2013vta, Banerjee:2013mca, Pimentel:2013gza,
Ghosh:2014kba, Larsen:2014wpa,Anninos:2014lwa,deHaro:2014xfa, Trivedi14}.  See also
other recent discussions in Refs.~\cite{Banks:2013qra, Banks:2013qpa,
Kiritsis:2013gia, Kol:2013msa, Kawai:2014vxa}.

While the connection between the bulk wavefunction and the CFT partition
function is intuitive and well-motivated, the absence of string theory
examples and a number of technical subtleties in the original
formulation of the duality initially cast doubt about its validity. 
A different route to this duality was put forward in Ref.~\cite{MS_HC09}. This was based 
on a fact that FRW spacetimes of a theory with potential $V$ are in
correspondence with domain-wall space times of a theory with potential
$-V$ \cite{Skenderis:2006jq, Skenderis:2006fb}. This correspondence
links accelerating FRW spacetimes to domain-wall space times, either
asymptotically AdS or asymptotic to non-conformal brane backgrounds,
where standard gauge/gravity duality is better understood. Computing
independently cosmological observables such as the spectra and
bi-spectra and correlation functions of the QFT dual to the
corresponding domain-wall spacetime  one could then analyse whether
there is any relation between the two. It turned out that in this way
one could establish precise holographic formulae for the spectra and
bi-spectra of general inflationary models in terms of a specific
analytic continuation of 2- and 3-point functions of the energy momentum
tensor of the dual QFT \cite{MS_HC09,MS_HC10,  MS_HCob10, MS_NG,
Easther:2011wh, MS_NGGW, BMS,McFadden:2013ria}. This provides an
effective holographic description of general single field inflationary model.

The holographic formulae relate cosmological in-in correlators with QFT
correlation functions of the energy momentum tensor. While these
formulae were obtained in the regime where gravity is valid, it was also
conjectured \cite{MS_NG,MS_NGGW} that the formulae hold more generally:
cosmological in-in correlators and the QFT correlators may receive
quantum corrections but their relation should  remain the same.  Indeed,
in a class of slow roll models one can establish the same formulae
without assuming the gravity approximation but by using that the
partition function of the dual QFT computes the wavefunction of the bulk
theory \cite{JYsingle}. 

This class of slow-roll models consists of single field models where the
bulk potential has two near-by de-Sitter critical points
\cite{BMS}. Correspondingly the dual QFT has two near-by fixed
points. One may use conformal perturbation theory to compute the QFT
correlation functions at strong coupling and in this way the slow-roll
results are exactly reproduced both for the spectra and bispectra  (to
the order checked) \cite{BMS,McFadden:2013ria}. For the same class of
models, one can alternatively compute the bulk wavefunction from the QFT
partition function by only assuming that  the dual QFT has  two near-by
fixed points \cite{JYsingle}, recovering the
holographic formulae derived in Refs.~\cite{MS_HC09, MS_NG, BMS}  (after taking into account the relation between the analytic continuations we discuss in appendix \ref{anal}) \footnote{More generally, there is agreement between explicit computations of 2- and 3-point functions computed using the wavefunction such as those in Refs.~\cite{Maldacena:2011mk,Mata:2012bx} and the formulae in
Refs.~\cite{MS_HC09, MS_NG, MS_NGGW} (after the analytic continuations
are matched, see appendix \ref{anal}, and modulo the semi-local
contributions which are often not computed using the wave function
approach).}.  

In more detail, the dual QFT was considered to be  a deformation of  a CFT by a
marginally relevant operator of dimension $3 -\lambda$, where $0<
\lambda \ll 1$ and the strength of the deformation is also proportional
to $\lambda$. Assuming that the coefficient in the 3-point function of
the deforming operator is positive and of order one (relative to
$\lambda$)  one may show that there is a nearby fixed point. If in
addition the UV  CFT contains no other nearly marginal operator the RG flow between the UV fixed point
to the IR fixed point is controlled by the deforming operator alone. In the
bulk this means that we are considering a single field model. 

The purpose of this paper is to relax some of these conditions.  While we
will still arrange for the theory to have two nearby fixed points we will
now assume that there are a multitude of deforming scalar operators.  Each of
these is related to a light bulk field and we now have a multi-field
inflationary model.   While recent observations appear to favour single field inflation~\cite{Ade:2013rta},  the presence of many nearly marginal operators is natural from the perspective of the dual QFT. 

Our main result is the computation of the power spectra for adiabatic
and entropy perturbations. We will be able to carry out this computation completely 
for a simple model. One of the main differences between single field and multi-field models 
is about the time dependence of the adiabatic curvature perturbation $\zeta$.
According to standard cosmological perturbation theory, the
adiabatic curvature perturbation $\zeta$ should be time independent at
large scales in single field models of inflation but it should not be conserved in the presence of 
entropy perturbations. It was indeed confirmed in Ref.~\cite{JYcsv} that the
holographically computed power spectrum for $\zeta$ is conserved in the
one field case (although some unresolved subtleties were pointed out
concerning the conservation of the bispectrum). In this paper we 
will demonstrate that the holographically computed spectra are not conserved due to the influence of the entropy
perturbation, as expected from cosmological perturbation theory (see for instance Ref.~\cite{GWBM}).

The outline of this paper is as follows. In
Sec.~\ref{Sec:Preliminaries} we review and extend the holographic setup for inflation to the multi-scalar case.
In Sec.~\ref{Sec:dCFT} we discuss the dual QFT and in
Sec.~\ref{Sec:RGcosmology} we relate the holographic RG flow with cosmological evolution 
in the bulk. In Sec.~\ref{Sec:primordial} we set up the computation of the spectra of primordial adiabatic and entropy 
perturbations using holography and we explicitly compute these spectra 
in a simple example in Sec.~\ref{Sec:example}. We present our concluding remarks in Sec.~\ref{Sec:conclusions}.
The paper also contains two appendices: in Appendix \ref{anal} we review
the analytic continuation from AdS to dS, which is a part of the
holographic dictionary while Appendix \ref{comp} contains an alternative
derivation of the power spectra.  

\section{Preliminaries}  \label{Sec:Preliminaries}
 
In this section we will use the approach developed in
Refs.~\cite{JYsingle, JYcsv} and extend the results of
Refs.~\cite{JYsingle, BMS} to the case of $A$ scalar fields. 

\subsection{Wave function}
The cosmological spacetime metric can be given in ADM formalism as
\begin{align}
 & \dd s^2 =  - N^2 \dd t^2 + h_{ij} \left( \dd x^i + N^i \dd t
 \right) \left( \dd x^j + N^j \dd t \right)\,. \label{Exp:metricb}
\end{align}
Here we shall restrict attention to the situation where the metric is
asymptotically de Sitter both in the far future and in the far past. 
I.e. we are describing the case of a period of slow roll inflation
followed by $\Lambda$ domination. 

Our starting point is the assumption that the wave function of
the bulk gravitational field is related to the generating functional of
a boundary QFT
\begin{align} 
 & \psi_{\rm bulk}[h,\phi] \propto Z_{{\rm QFT}_+}[h,\phi]\,, \label{duality}
 \end{align}
where the generating functional $Z_{{\rm QFT}_+}$ is given by
\begin{equation}
Z_{{\rm QFT}_+} [h,\phi]  =  e^{-W_{{\rm QFT}_+}[h,\phi]} =  \int D \chi\,
\exp \left(  - S_{{\rm QFT}_+} [\chi,\, h,\, \phi]\right)   \, \label{Exp:Z}.
\end{equation}
Here $\psi_{\rm bulk}$ denotes the wave function of the bulk and $\chi$ denotes boundary
fields. The metric $h_{ij}$ and the inflaton field (or fields) $\phi$ act as
sources for corresponding gauge invariant operators constructed using
$\chi$ (we will often suppress the indices of $h_{ij}$). 
$Z_{\rm QFT_+} [h,\phi]$ is understood as the collection of
correlation functions of gauge invariant operators. The QFT should admit a 't Hooft large $N$ limit and 
is taken to be Euclidean (though this is not essential -- the Euclidean
theory may be related by Wick rotation to a Lorentzian QFT). 

To date there is no string/M theory derivation of the dual QFT 
and there are questions whether such a theory exists non-perturbatively (see, for example, Refs.~\cite{Dyson:2002nt, Goheer:2002vf} for 
early discussions of this point in the context of dS/CFT). However,  one
can give an operational definition of the theory in large $N$
perturbation theory and  infer many its properties (within perturbation
theory). In Ref.~\cite{Maldacena02} Maldacena used the bulk description in
order to extract correlators of the putative dual CFT${}_+$ of de Sitter
and he showed the these correlators are related by analytic continuation
to the correlation functions of the CFT${}_-$ dual to AdS \footnote{Here
and below we will use the subscripts + and - to refer to the QFT dual  
of the cosmology and domain-wall space-time, respectively. These
spacetimes  are linked via the domain-wall/cosmology correspondence
\cite{Skenderis:2006jq, Skenderis:2006fb}.},  see appendix \ref{anal} for a review of the analytic continuation.
In Refs.~\cite{MS_HC09, MS_NG, MS_NGGW}  it was shown that correlation functions of QFT${}_+$  (the putative dual to general single field inflationary spacetimes)  are related by analytic
continuation to correlation functions  of QFT${}_-$ (the QFT dual to the corresponding domain-wall space-time \cite{Skenderis:2006jq, Skenderis:2006fb}) in the large $N$ limit.  
QFT${}_+$ was dubbed "pseudo-QFT" in Ref.~\cite{MS_HC09} since it was only defined via analytic continuation from QFT${}_-$.  A duality  where the CFT${}_+$ has an a priori definition was proposed in Ref.~\cite{Anninos:2011ui}. In this duality the CFT was conjectured  to be dual to the dS solution of Vasiliev's higher spin gravity rather than Einstein gravity. At the perturbative level the CFT${}_+$ is related by analytic continuation 
to the CFT${}_-$ dual to AdS, as in Ref.~\cite{MS_HC09}, but now one could analyse the theory non-perturbatively and such analyses were presented in Refs.~\cite{Anninos:2012ft, Banerjee:2013mca}. 

Since the analytic continuation takes $N \to -N$, it is unclear whether
the QFT${}_+$ and QFT${}_-$ are related to each other by this analytic
continuation non-perturbatively in $1/N$.  A recent study
\cite{Codesido:2014oua} of a specific (highly supersymmetric) example showed that the partition
function of QFT${}_-$ can be extended to the entire complex plane as
a function $N$. However, the non-perturbative contributions become
significant as $N \to -N$ and may potentially upset the agreement
between the holographic and cosmological computations. On the other
hand, one may have theories that coincide in large $N$ perturbation
theory and only differ non-perturbatively, as was recently shown in
Ref.~\cite{Vafa:2014iua}. Thus, the dual to cosmology may only agree
with the analytic continuation of QFT${}_-$ in large $N$ perturbation
theory. 

In this paper we will set aside such non-perturbative issues and focus only on the leading contribution in the large $N$ limit.
We will do computations in QFT${}_-$ and then analytically continue to obtain the correlators relevant for cosmology.

\subsection{Correlators in the bulk}
Once we are given the wave function $\psi_{\rm bulk}$, we can compute the
correlators for the bulk. We describe the perturbation of the spatial metric as
\begin{align}
 & h_{ij} = a^2(t) e^{2\zeta(t,\, \sbm{x})} \delta_{ij}\,,  \label{Exp:gauge}
\end{align}
where we neglected transverse traceless perturbations (gravitational waves). We will specify the
time slicing later. In this paper, extending the analysis
in Ref.~\cite{JYsingle}, we consider the case of $A$ scalar fields in
the bulk. Then, in addition to the curvature perturbation $\zeta$, there are
$(A-1)$ scalar degrees of freedom, $s^{a'}$ with
$a' = 2,\, \cdots, A$, the entropy perturbations. In
the following, we collect together the curvature perturbation $\zeta$ and the
entropy perturbation $s^{a'}$ in $\Phi^a$ ($a = 1,\, \ldots,\, A$), where
$\Phi^1(t,\,\bm{x})\equiv \zeta(t,\, \bm{x})$ and
$\Phi^{a'}(t,\, \bm{x})\equiv s^{a'}(t,\, \bm{x})$. The index $a$ is
raised and lowered by the Kronecker delta $\delta_{ab}$.

As in the discussion in the previous subsection, the wave function in
the bulk is related to the generating functional of the connected
correlators of gauge invariant operators as 
\begin{align}
 & \psi_{\rm bulk}[\Phi^a] = {\cal N}\, Z_{{\rm QFT}_+}[\Phi^a] =
 {\cal N}\, e^{-  \WQ[\Phi^a]}\,, \label{Exp:psi}
\end{align}
with
\begin{align}
 & \WQ[\Phi^a] \equiv - \ln Z_{{\rm QFT}_+}[\Phi^a] \,,
\end{align}
where ${\cal N}$  is a normalisation constant. Using the wave function $\psi_{\rm bulk}[\Phi^a]$, the probability density function
$P[\Phi^a]$ is given by  
\begin{align}
 P[\Phi^a] = \left| \psi[\Phi^a] \right|^2
 = |{\cal N}|^2\, e^{- 2 {\rm Re} \left[\WQ[\Phi^a]  \right] }\,.
\end{align}
Once we have the probability density $P[\Phi^a]$, we can calculate cosmological $n$-point
functions as
\begin{align} 
 & \langle \Phi^{a_1}(\bm{x}_1) \Phi^{a_2}(\bm{x}_2)  \cdots
 \Phi^{a_n}(\bm{x}_n) \rangle = \int  D \Phi
  \, P[\Phi^a]\, \Phi^{a_1}(\bm{x}_1) \Phi^{a_2}(\bm{x}_2)  \cdots
 \Phi^{a_n}(\bm{x}_n)\,, \label{npt}
\end{align}
where $D \Phi$ denotes the integration measures for $\zeta$ and
$s$, given by
\begin{align}
 & D \Phi \equiv D\zeta \prod_{a'=1}^{A-1} D s^{a'}\,.
\end{align}
The integration measure is not part of  the boundary QFT data since the
fields $\Phi^a$ are external fields and hence some
additional input may be necessary. Changes in the measure may be 
represented by local terms in the integrand, which may be incorporated
by a redefinition of $W_{\rm QFT}$. Here we will assume that there are no non-trivial measure factors. In principle, one can determine 
the measure factors in the semi-classical regime using the derivation of
the holographic formulae via the domain-wall/cosmology correspondence.
In the single field case this indeed yields a trivial measure.

Eliminating the background contribution $\WQ[\Phi^a=0]$ by a
redefinition of ${\cal N}$, the probability density function $P[\Phi^a]$ is given by
\begin{align}
 & P[\Phi^a]=  |{\cal N}|^2  e^{- \delta W[\Phi^a] } \,,
\end{align}
where we defined
\begin{align}
 & \delta W[\Phi^a] \equiv 2 {\rm Re} \left[ \WQ[\Phi^a] -
 \WQ[\Phi^a=0]  \right]\,.
\end{align}
Assuming that the wave function is normalizable,  we then fix the normalization constant ${\cal N}$ by
\begin{align}
 & \int D \Phi \, P[\Phi^a] =1\,. \label{Cond:norm}
\end{align}

We now expand $\delta W[\Phi^a]$ as
\begin{align}
 & \delta W[\Phi^a] = \sum_{n=1}^A \frac{1}{n!} \int \dd^d \bm{x}_1
 \cdots \int \dd^d \bm{x}_n  W^{(n)}_{a_1 \cdots a_n} (\bm{x}_1,\, \cdots,\,
 \bm{x}_n ) \Phi^{a_1}(\bm{x}_1) \cdots \Phi^{a_n} (\bm{x}_n) \,, \label{Exp:dWn}
\end{align}
where
\begin{align}
 & W^{(n)}_{a_1 \cdots a_n} (\bm{x}_1,\, \cdots,\,
 \bm{x}_n ) \equiv 2 {\rm Re} \left[ \frac{\delta^n \WQ[\Phi^a]}{\delta
 \Phi^{a_1}(\bm{x}_1) \cdots \delta \Phi^{a_n}(\bm{x}_n) }
 \Bigg|_{\Phi^a=0} \right]\,. \label{Def:Wn}
\end{align}
Once we obtain
$W^{(n)}_{a_1 \cdots a_n} (\bm{x}_1,\, \cdots,\, \bm{x}_n)$, we can compute
the $n$-point functions using Feynman rules~\cite{JYsingle}. In more detail, 
we view (\ref{Exp:dWn}) as our action with $\Phi^a$ treated as elementary fields. Then 
the inverse of the quadratic part 
provides the propagator for the Feynman rules while the higher order terms provide the vertices.

For example, the two-point function on $\Phi^a(\bm{x})$ is given by
\begin{align}
 & \langle \Phi_{a_1} (\bm{x}_1) \Phi_{a_2} (\bm{x}_2) \rangle = W_{a_1
 a_2}^{(2)\, -1} (\bm{x}_1,\, \bm{x}_2) \,, \label{Exp:PP}
\end{align}
where $W_{a_1 a_2}^{(2)\, -1} (\bm{x}_1,\, \bm{x}_2)$ denotes
the inverse matrix
of $W_{a_1 a_2}^{(2)}(\bm{x}_1,\, \bm{x}_2)$, 
\begin{align}
 & \sum_{a_2=1}^A \int \dd^d \bm{y} W_{a_1 a_2 }^{(2)} (\bm{x},\, \bm{y})
 W_{a_2 a_3}^{(2)\, -1} (\bm{y},\, \bm{z}) = \delta_{a_1 a_3} \delta(\bm{x}-
 \bm{z})\,.  \label{Rel:W2W2I}
\end{align}
Furthermore, $W_{a_1 a_2}^{(2)}(\bm{x}_1,\, \bm{x}_2)$ equal minus the
two-point function of the operators ${\cal O}_{\Phi^a}$ dual to scalar $\Phi^a$ 
\begin{align}
 & W_{a_1 a_2}^{(2)}(\bm{x}_1,\, \bm{x}_2) = - 2 {\rm Re} \left[ \langle
 {\cal O}_{\Phi^{a_1}}(\bm{x}_1) {\cal O}_{\Phi^{a_2}} (\bm{x}_2) \rangle_+  \right] \,,  \label{2pt}
\end{align}
where $\langle \  \rangle_+$ indicates correlators of
QFT${}_+$. (In the following sections we will often change the
basis of operators we consider. In such cases one needs to  take into
account the Jacobian of the transformation to compute correctly the
2-point functions.) 

As reviewed in the previous subsection, we are currently lacking a proper definition of QFT${}_+$. However, we know that in the large $N$ limit we can compute the correlators in QFT${}_-$ (which is a standard 
QFT) and then analytically continue the result according to the rules discussed in appendix \ref{anal}. All QFT computation 
in this paper are done in standard QFT and they are about QFT${}_-$. To obtain the final cosmological results proper care 
was taken to implement the necessary analytic continuation. Note that
the derivation of the relation between bulk and boundary correlators
given in Sec.~\ref{Sec:primordial} applies irrespective of how QFT$_+$
is provided. In the following sections to simplify the presentation we
will suppress the distinction between QFT${}_-$ and QFT${}_+$ and simply
refer to the dual QFT. Hopefully this will not cause any confusion.

\section{Dual quantum field theory}   \label{Sec:dCFT}
In this section, we discuss the $d$-dimensional quantum field theory which would be
dual to the $(d+1)$-dimensional inflationary spacetime. 
We shall assume that $d$ is odd to avoid complications due to the gravitational conformal anomalies (our main interest is four bulk dimensions, so $d=3$).  Since we are considering space-times that are asymptotically de Sitter,
the dual theory is a deformation of a CFT,
\begin{align}
 & S_{\rm QFT} [\chi] = S_{\rm CFT} [\chi] + \sum_{a=1}^A \int \dd \Omega_d u^a
 O_a(\bm{x}) \,, \label{Exp:Su}
\end{align}
where $\dd \Omega_d$ is the $d$-dimensional invariant volume and 
$S_{\rm CFT}$ is the action at the UV fixed point\footnote{The dual QFT
takes a similar form also near the IR fixed point.}. This CFT is
deformed by gauge invariant operators $O_a(\bm{x})$ of dimension $\Delta_a<d$ and 
$u^a$ are the corresponding coupling constants. Hereafter, we suppress the
summation over $a$. In this paper, we assume that the
field space for $u^a$ is flat and hence we raise and lower indices $a$ by
the Kronecker delta. 
In this section, we solve the RG flow in the flat space and determine
the renormalisation scale dependence of the coupling constants $u^a$.

\subsection{RG flow} \label{Sec:RG}

The relevant operators in Eq.~(\ref{Exp:Su})  induce an RG flow which we will study in this section.
As long as the deformation parameters are small one can study the RG flow using conformal perturbation theory.
To prepare for this computation let us  summarise standard results for a CFT on
flat $\mathbb{R}^d$. Conformal symmetry
determines the two-point function and the three point function up to
constant parameters as
\begin{align}
 & \langle O_a(\bm{x}) O_b(\bm{y}) \rangle_{\rm CFT} = \frac{c_a
 \delta_{ab}}{|\bm{x} - \bm{y}|^{2\Delta_a}} \,,  \label{Eq:2pCFT}
\end{align}
and
\begin{align}
 & \langle O_a(\bm{x}) O_b(\bm{y}) O_c(\bm{z}) \rangle_{\rm CFT} =
 \frac{C_{abc}}{|\bm{x} - \bm{y}|^{\Delta_a + \Delta_b - \Delta_c}
 |\bm{y} - \bm{z}|^{\Delta_b + \Delta_c - \Delta_a} |\bm{z} -
 \bm{x}|^{\Delta_c + \Delta_a - \Delta_b}} \,,  \label{Eq:3pCFT}
\end{align}
where $\Delta_a$ is the scaling dimension of the operator $O_a$. These imply the 
operator product expansion (OPE) 
\begin{align}
 & O_a(\bm{x}) O_b(\bm{y}) = \frac{c_a \delta_{ab}}{|\bm{x} -
 \bm{y}|^{2\Delta_a}} + \delta^{cd}\, \frac{C_{abd}}{c_c} \frac{O_c(\bm{x})}{|\bm{x} -
 \bm{y}|^{\Delta_a + \Delta_b - \Delta_c}} + \cdots \,,  \label{OPE}
\end{align}
 where the dots denote less singular terms as $\bm{x} \to \bm{y}$.

Following Ref.~\cite{Klebanov11}, we now study the RG flow for the theory with
action (\ref{Exp:Su}). The generating function for the QFT is given by
\begin{align}
 & Z_{\rm QFT} = \int D \chi
\exp \left( - S_{\rm CFT} - \int \dd^d \bm{x} u^a O_a(\bm{x}) \right)\,. \label{Exp:ZdCFT}
\end{align}
Let us introduce a UV cutoff $\mu_0$. In the regulated theory the correlators are given by
\begin{align}
 & \langle O_{a_1}(\bm{x}_1) \cdots O_{a_n}(\bm{x}_n) \rangle_{\mu_0}
  \cr
&\,= \frac{1}{Z_{\rm QFT}} \int D \chi  O_{a_1}(\bm{x}_1) \cdots
 O_{a_n}(\bm{x}_n) \exp \left( - S_{\rm CFT} - \int \dd^d \bm{x} u^a_0
 O_a(\bm{x}) \right) \,, \label{Exp:On}
\end{align}
where we denote the coupling constants $u^a$ at $\mu=\mu_0$ by
\begin{align}
 & u^a_0 \equiv u^a(\mu_0)\,, 
\end{align}
and we consider all points 
$\bm{x}_i$ ($i=1, \ldots, n$) separated from each other by 
distance more than $1/\mu_0$. We can now express the correlation functions of the QFT in terms of
CFT correlators with the cutoff by expanding the
exponential factor in the generating functional as
\begin{align}
 & e^{- S_{\rm CFT}- \int \dd^d \sbm{x}\, u^a_0 O_a(\sbm{x})} \cr
 &= \left[ 1 - \int \dd^d
 \bm{x}\, u^a_0 O_a(\bm{x})  + \frac{1}{2}  \int \dd^d
 \bm{x}\,\int_{|\sbm{x} - \sbm{y}| > 1/\mu_0} \hspace{-2pt} \dd^d \bm{y}\,  u^a_0
 u^b_0\, O_a(\bm{x}) O_b(\bm{y}) + \cdots  \right] e^{- S_{\rm CFT}} \,.
 \cr \label{expansion}
\end{align}

Let us now integrate out the modes between $\mu$ and
$\mu_0$. Using the OPE (\ref{OPE}) in the third term of
Eq.~(\ref{expansion}) we obtain
\begin{align}
 &  \frac{1}{2}  \int \dd^d \bm{x}\,\int_{1/\mu_0 <|\sbm{x} - \sbm{y}| <
 1/\mu} \hspace{-2pt} \dd^d \bm{y}\,  u^a_0  u^b_0\, O_a(\bm{x})
 O_b(\bm{y}) \cr
 & = \frac{1}{2} u^b_0 u^c_0\, \delta^{ad}\,
 \frac{C_{bcd}}{c_a}   \int \dd^d \bm{x}
 O_a(\bm{x}) \int_{1/\mu_0 <|\sbm{x} - \sbm{y}| < 1/\mu} \hspace{-2pt}
 \dd^d \bm{y}\, \frac{1}{|\bm{x} - \bm{y}|^{\Delta_b(\mu_0) + \Delta_{c}(\mu_0) -
 \Delta_{a}(\mu_0)}} + \cdots \cr
 & =  -  \frac{1}{2} u^b_0 u^c_0 \tilde{C}^a_{bc}
 \frac{\mu^{\lambda_a- \lambda_b - \lambda_c} - \mu_0^{\lambda_a -
 \lambda_b - \lambda_c}}{\lambda_a -\lambda_b - \lambda_c} \int \dd^d \bm{x}
 O_a(\bm{x}) + \cdots  \,, \label{integration}
\end{align}
where
\begin{align}
 & \lambda_a \equiv d-\Delta_a(\mu_0) \,.  \label{Def:lambdaa}
\end{align}
Here we introduced
\begin{align}
 & \tilde{C}^a_{cb} \equiv {\rm Vol}(S^{d-1})\, \delta^{ad}\,
 \frac{C_{bcd}}{c_a} = \frac{2 \pi^{d/2}}{\Gamma(d/2)}  \delta^{ad}\,
 \frac{C_{bcd}}{c_a} \,,
\end{align}
where $ {\rm Vol}(S^{d-1})$ is the volume of the round $(d-1)$-dimensional
sphere. The integration of the UV modes gives rise to the running
of the coupling constant $u^a$ as
\begin{align}
 & u^a(\mu) = u^a_0 +  \frac{1}{2} u^b_0 u^c_0 \tilde{C}^a_{bc}\,
 \frac{\mu^{\lambda_a - \lambda_b - \lambda_c} -
 \mu_0^{\lambda_a - \lambda_b - \lambda_c}}{\lambda_a -
 \lambda_b - \lambda_c} + {\cal O} (u_0^3)
 \,. \label{Exp:uarun}
\end{align}
In the second and third lines of Eq.~(\ref{integration}), we included
only the terms which contribute to the running of the coupling constants
$u^a$.

Let us now introduce the dimensionless coupling constants $g^a$ as
\begin{align}
 & g^a(\mu) \equiv \mu^{-\lambda_a} u^a (\mu) \,.
\end{align}
By using Eq.~(\ref{Exp:uarun}), the running of $g^a(\mu)$ is given by
\begin{align}
 & g^a(\mu) = \left( \mu \over \mu_0 \right)^{-\lambda_a} g^a_0  +
 \frac{1}{2} g^b_0 g^c_0 \tilde{C}^a_{bc}
 \frac{  \left( \mu \over \mu_0 \right)^{-(\lambda_b + \lambda_c)} -
  \left(\mu \over \mu_0 \right)^{-\lambda_a}}{\lambda_a -
 \lambda_b - \lambda_c} + {\cal
 O} (g_0^3)  \,,  \label{Exp:ga}
\end{align}
where $g^a_0 \equiv \mu_0^{-\lambda_a} u^a (\mu_0)$. 
The beta function 
\begin{align}
 & \beta^a(\mu) \equiv \frac{\dd g^a(\mu)}{\dd \ln \mu} \label{Def:beta}
\end{align}
can now be computed by
inserting Eq.~(\ref{Exp:ga}) into Eq.~(\ref{Def:beta}), yielding
\begin{align}
 & \beta^a(\mu) = -\lambda_a\, g^a(\mu) + \frac{1}{2} g^b(\mu) g^c(\mu)
 \tilde{C}^a_{bc}  + {\cal O} (g^3)  \label{Eq:beta}
\end{align}
(there is no sum over $a$ in the first term and there are implicit sums over $b$ and $c$ in the second term).
The second term stems from the new short distance singularities due to the composite operators 
and this leads to the
deviation from the classical scaling. This analysis is valid as long as
$g^a$ is small and hence we can solve the RG flow
perturbatively. Note that the beta function does not depend on the UV
cutoff $\mu_0$ explicitly, so we can send it to the infinity.

\subsection{Separable case}  \label{SSec:multig}

Having obtained the beta functions one may now investigate the dynamics along the RG flow and possible IR fixed points.
For cases where $g^a$ are small  throughout the RG flow from the UV to
the IR fixed point one may use conformal perturbation theory to compute
the correlators and from those extract the cosmological prediction.  
The dynamics in such cases is essentially controlled by the form of the
OPE. Leaving a general study for future work, in this paper we focus on
the simplest possibility: the $A$ operators do not couple to each other
at leading order, i.e. three point function involving more than one type
of operator are zero. This implies that the structure constants
$C_{abc}$ and $\tilde{C}^a_{bc}$ satisfy 
\begin{align}
 & C_{abc} =C_a \delta_{ab} \delta_{bc}\,, \qquad
  \tilde{C}^a_{bc} = \tilde{C}_a \delta^{ac} \delta_{bc}
  \,,  \label{Cond:diagonal}
\end{align}
where $\tilde{C}_a = (2 \pi^{d/2}/\Gamma(d/2)) (C_a/c_a)$
(no sum over $a$). 
In this case, the RG equation becomes separable as
\begin{align}
 & \beta^a(\mu)= -\lambda_a g^a(\mu) + \frac{1}{2}
 \tilde{C}_a \left(g^a(\mu)\right)^2 + {\cal O} (g^3)\,. \label{Eq:betam}
\end{align}
We essentially have $A$ decoupled copies of the case of a single deforming  operator.
One may thus directly borrow the results obtained for this case
\cite{BMS, McFadden:2013ria, Klebanov11}.

Assuming $g^a> 0$ and $\tilde{C}_a >0$ we find that there is an IR fixed point at 
\begin{equation}
g^a_*= \frac{2 \lambda_a}{\tilde{C}_a}.
\end{equation} 
In this case, Eq.~(\ref{Eq:betam}) can be solved as
\begin{align}
 & g^a(\mu) =  \frac{2 g^a(\mu_a) }{1+ \left( \mu/\mu_a
 \right)^{\lambda_a}} , \qquad g^a(\mu_a) \equiv
 \frac{\lambda_a}{\tilde{C}_a}\,, \label{Sol:gs} 
\end{align}
where a pivot scale $\mu_a$ is introduced as an integration constant. For the coupling to remain small throughout the RG flow we need 
$\lambda_a/\tilde{C}_a \ll 1$. In the UV, $\mu \gg \mu_a$, $g^a(\mu) \to
0$, while in the IR, $\mu \ll \mu_a$, $g^a(\mu)  \to g^a_*=2
g^a(\mu_a)$.

\section{RG flow and cosmological evolution}  \label{Sec:RGcosmology}
In the previous section, we derived the RG equation for the coupling
constant $g^a$. In this section, we discuss the
bulk evolution of the corresponding bulk scalar fields $\phi^a$.

\subsection{Superpotential}

We consider gravity coupled to $A$ scalar fields governed by the action
\begin{equation}
S = \frac{1}{2 \kappa^2} \int \dd^{d+1} x \sqrt{-g} \left[R - (\partial \phi^a)^2  - 2 \kappa^2 V(\phi^a) \right].
\end{equation}
The scalar field equations in a $(d+1)$-dimensional FRW universe are given by
\begin{align}
 & \ddot{\phi}^a + d H \dot{\phi}^a + \kappa^2 \frac{\partial V(\aph)}{\partial \phi^a}
 = 0\,, \label{Eq:KG}
\end{align}
where $H$ is the Hubble parameter, which satisfies
\begin{align}
 & H^2 = \frac{1}{d(d-1)} \left[ \sum_{a=1}^A (\dot{\phi}^a)^2 +
 2 \kappa^2 V(\aph) \right]\,, \label{Eq:Friedmann}
\end{align}
where $\kappa^2 \equiv 8 \pi G$ is the gravitational constant. Here,
$\aph$ denotes the set of variables $\phi^a$ ($a=1, \ldots, A$).

One may trade (under some assumptions) the second order field equations (\ref{Eq:KG})-(\ref{Eq:Friedmann}) 
by first order equation:
\begin{align}
 & \dot{\phi}^a =  \frac{\partial
 W(\aph)}{\partial \phi^a} \,, \label{Eq:dphi} \\
 & H = -\frac{1}{d-1} W(\aph)\,, \label{Eq:HW}
\end{align}
where $W(\aph)$ is the so-called superpotential, which is related to the
potential $V(\aph)$ by
\begin{align}
 & 2 \kappa^2 V(\aph) = \frac{d}{d-1}
 W^2(\aph) - \sum_{a=1}^A \left( \frac{\partial
 W(\aph)}{\partial \phi^a} \right)^2 \,. \label{Exp:VW}
\end{align}
This reformulation was first obtained in cosmology using Hamilton-Jacobi theory in 
Ref.~\cite{Salopek:1990jq}.  It was rediscovered in gauge/gravity duality in
Refs.~\cite{Skenderis:2006jq, Skenderis:1999mm, DeWolfe:1999cp,
Freedman:2003ax} where it was linked with (fake) supersymmetry and gravitational stability.

One can verify that any solution of the first order equations also solves the second order equations
(\ref{Eq:KG}) and Eq.~(\ref{Eq:Friedmann}). Indeed taking the partial
derivative of Eq.~(\ref{Exp:VW}) with respect to $\phi^a$ we obtain
\begin{align*}
 & \kappa^2 \frac{\partial   V(\aph)}{\partial \phi^a} = 
 \frac{d}{d-1} W(\aph) \frac{\partial W(\aph)}{\partial \phi^a} -
 \sum_{a=1}^A \sum_{b=1}^A  \frac{\partial
 W(\aph)}{\partial \phi^b}  \frac{\partial^2
 W(\aph)}{\partial \phi^a \partial \phi^b}
\,.
\end{align*}
Using Eqs.~(\ref{Eq:dphi}) and (\ref{Eq:HW}) we find that the first
term of the right hand side is $-dH\dot{\phi}^a$ and the second term is
$- \ddot{\phi}^a$. Meanwhile, Eq.~(\ref{Exp:VW}) with Eq.~(\ref{Eq:HW}) is nothing but the
Friedmann equation (\ref{Eq:Friedmann}).  
The converse (i.e. going from Eqs.~(\ref{Eq:KG})-(\ref{Eq:Friedmann}) to Eqs.~(\ref{Eq:dphi})-(\ref{Eq:HW})-(\ref{Exp:VW}))
is also true but under some additional assumptions, see Ref.~\cite{Skenderis:2006jq}.

\subsection{Potential for the solvable RG flow}

We would now like to link the field theory discussion with bulk
dynamics. The specific example we are considering is $A$ decoupled
copies of the single field case studied in Refs.~\cite{JYsingle,
BMS,McFadden:2013ria,JYcsv} so one may borrow many of the results we
need for our case.  The properties of the potential near the critical
point are fixed from those of the CFT.  In particular, the spectrum of
operators and the OPE coefficients fix the form of the potential through
cubic order in the fields \cite{BMS}. We actually know the QFT along the
entire RG flow as an expansion in the $\lambda_a$. This fixes the
superpotential (and thus the potential) as an expansion in $\lambda_a$.
It was shown in Ref.~\cite{McFadden:2013ria} that in the single field case
the superpotential in $d=3$ (and in our notation) is given by 
\begin{equation}
W(\phi) = -2 \exp \left( -\frac{1}{2} \int_0^\phi \dd g \beta(g) \right) +
 {\cal O}(\lambda^7)\,. 
\end{equation}
This is equivalent to saying that there is a QFT scheme such that \cite{McFadden:2013ria}
\begin{equation} \label{beta_W}
-2 W'(\phi)/W(\phi) = \beta(\phi) + {\cal O}(\lambda^6) \,. 
\end{equation}
One should compare this expression with the proposal \cite{de Boer:1999xf} that the QFT beta function is given exactly by (now for general $d$ and many couplings)
\begin{equation} \label{beta}
\beta^a = - (d-1) \frac{\partial}{\partial \phi^a} \ln
 W(\phi^a)\, .
\end{equation}
This scheme was also the choice used in early works on holographic cosmology \cite{Larsen:2002et,LM03,vdS}. 
One should note however that in cases where the RG flow is driven by supersymmetric operators, Eq.~(\ref{beta}) is not the correct QFT beta function \cite{Bianchi:2001de,Bianchi:2001kw}, so one should use Eq.~(\ref{beta}) with caution.

Standard holographic methods compute renormalised correlators \cite{Skenderis:2002wp}.  On the cosmology side
we correspondingly obtain the cosmological observables at the end of the
inflationary period ($t \to \infty$). Here however we would like to
study one of the main properties that distinguish single vs multi-scalar
inflation, namely whether the perturbations are conserved outside the
horizon or not. To do this we need to understand the time dependence of
the cosmological perturbations. Following general expectations, we assume
that the coupling constant $g^a$ is related to the
scalar field $\phi^a$ as
\begin{align}
 & g^a(\mu,\, \bm{x}) =  \phi^a(t(\mu),\, \bm{x}) \,, \label{Rel:gphi}
\end{align}
and the renormalization scale is related to the scale factor by
\begin{align}
 & \frac{\mu}{\mu_0} = \frac{a}{a_0}. \label{Rel:mua}
\end{align}
(See also Refs.~\cite{BMS, JYcsv, AJ09, Alex11}.) With this choice, the
power spectrum of the curvature perturbation $\zeta$ is conserved for single field
model~\cite{JYsingle}.  One should note however that this prescription does not produce conserved bispectrum.
This may be related with the well-known difficulties with realising Wilsonian RG using holography \cite{Heemskerk:2010hk}.
In any case, in this paper we only discuss the power spectra and for those the prescription is adequate.
As discussed in
Ref.~\cite{JYcsv}, changing the linear relation (\ref{Rel:gphi}) to a
more general (local) function $g^a[\aph]$ can be understood simply as a
field redefinition. 

The scheme (\ref{Rel:gphi})-(\ref{Rel:mua}) implies (\ref{beta}).  Indeed,
using Eqs.~(\ref{Rel:gphi}) and (\ref{Rel:mua}), we obtain
\begin{align}
 & \beta^a =  \frac{\dot{\phi}^a}{H} = -
 (d-1) \frac{\partial}{\partial \phi^a} \ln
 W(\aph)\,. \label{Rel:betaW}
\end{align}
Inserting this expression into Eq.~(\ref{Exp:VW}), we can compute the potential as
\begin{align}
 & 2 \kappa^2 V(\aph) = \left(\frac{W(\aph)}{d-1}\right)^2 \left[ d(d-1)
  - \sum_{a=1}^A (\beta^a)^2  \right]\,. \label{Exp:VW2}
\end{align}

\begin{figure}[t]
\begin{center}
\includegraphics[scale=0.6]{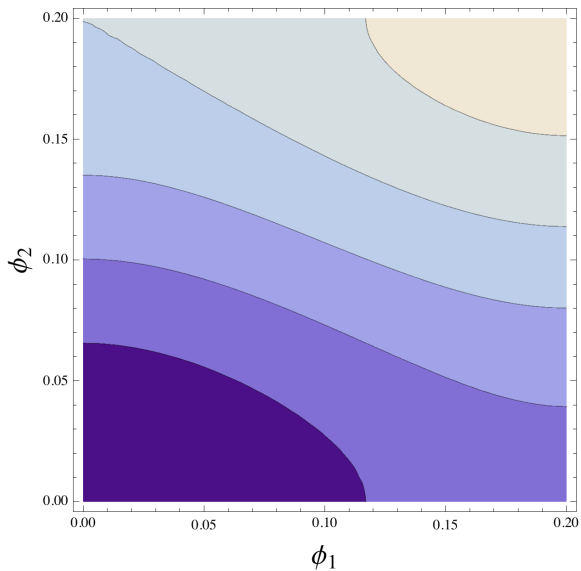}
\end{center}
\caption{Contour plot of the potential for the
 two-field case with $\lambda_1=0.02$, $\tilde{C}_1=0.2$,
 $\lambda_2=0.05$, $\tilde{C}_2=0.5$, $d=3$. In this
 case, $\phi_1$ and $\phi_2$ vanish in the UV fixed point and they take the same
 value $0.2$ in the IR fixed point. The brighter region shows the range
of $\phi_1$ and $\phi_2$ with larger values of
 $V(\phi_1,\, \phi_2)$. }
\label{Fg:HVforN}
\end{figure}

In our example in Sec.~\ref{SSec:multig}, the RG equation
for $g^a$ is separable, see  Eq.~(\ref{Eq:betam}). In this case, the corresponding superpotential
takes the form
\begin{align} \label{WWa}
 & W(\aph)=  W_0 \prod_{a=1}^A W_a(\phi^a)\,,
\end{align}
 where $W_a$ is a function of $\phi^a$ only. It follows from Eq.~(\ref{beta})
\begin{equation}
 W_a(\phi^a)= \exp\left[ - \frac{1}{d-1} \int \dd
 g^a \beta^a(g^a) \right] \,, \label{Exp:Wex}
\end{equation}
where $W_0$ in Eq.~(\ref{WWa}) is a constant that originates from the
integration constants when we integrate the beta functions in
Eq.~(\ref{Exp:Wex}). Using Eqs.~(\ref{Eq:HW}), (\ref{Exp:VW2}), (\ref{WWa}) and
(\ref{Exp:Wex}), we obtain the Hubble parameter $H(\aph)$ as
\begin{align}
 & H(\aph)= -\frac{W_0}{d-1}  \prod_{a=1}^A \exp\left[ - \frac{1}{2 (d-1)} \left(
 -\lambda_a (\phi^a)^2 + \frac{\tilde{C}_a}{3} (\phi^a)^3 + {\cal O}((\phi^a)^4)) \right)  \right]\,,
\end{align}
which implies that $W_0$ determines the de Sitter radius of the dS
critical point at $\phi^a=0$. The potential $V(\aph)$ is now equal to 
\begin{align}
 V(\aph) &= \frac{H^2(\aph)}{2\kappa^2} \left[ d(d-1)
  - \sum_{a=1}^A \left( -\lambda_a \phi^a + \frac{\tilde{C}_a}{2}
 (\phi^a)^2 + {\cal O}\left((\phi^a)^3 \right)\right)^2  \right]\,. \label{Exp:Vphi}
\end{align}
While the beta functions $\beta^a$ are in the separable form for each $a$, the Klein-Gordon equation for $\phi^a$ is
not separable because $\partial V(\aph)/\partial \phi^a$
also contains $\phi^b$ with $b \neq a$.

To get an intuition about the potential, we show in Fig.~\ref{Fg:HVforN}
the contour plot of  the potential for the two-field case, $V(\phi_1,
\phi_2)$, with $\lambda_1=0.02$ and $\lambda_2=0.05$. Brighter regions
denote larger values of the potential. 

\section{Primordial perturbations} \label{Sec:primordial}
In this section we introduce adiabatic and entropy
perturbations and we derive the relation between
$W^{(2)}_{ab}(\bm{x}_1,\,\bm{x}_2)$ and the two point functions of the
boundary operators.

\subsection{Boundary QFT}
In this section, we further study the case of $A$ deformation operators,
\begin{align}
& S_{\rm QFT}[g^a,\,\chi] = S_{\rm CFT}
 + \sum_{a=1}^A \int \dd \Omega\, g^a(\mu,\,\bm{x})
 {\cal O}_a(\mu,\, \bm{x})\,,  \label{QFT_action}
 \end{align}
where we introduced 
\begin{align}
 & g^a(\mu,\, \bm{x})\equiv \mu^{-\lambda_a}u^a(\mu,\, \bm{x})\,, \\
 & {\cal O}_a(\mu,\,\bm{x}) \equiv \mu^{\lambda_a} O_a(\mu,\,\bm{x})\,.
\end{align}
We consider couplings  $g^a(\mu,\,\bm{x})$ of the form
\begin{align}
 & g^a(\mu,\,\bm{x}) = g^a(\mu) + \delta g^a(\mu,\,\bm{x}) \,,
\end{align}
where $g^a(\mu)$ is the background value, corresponding to the FRW as described in the previous section. 
In the following only the background value of $g^a(\mu)$ and the
fluctuation $\delta g^a(\mu,\, \bm{x})$ appear, so to simplify the notation we will suppress the arguments and (with abuse of notation)
will denote the background value by $g^a$.

\subsection{Adiabatic and entropy perturbations in cosmological perturbation theory}
We decompose the fluctuations of the scalar field
$\delta \phi^a(t,\, \bm{x})$ into a fluctuation $\delta \sigma$  tangent to the background trajectory (adiabatic perturbations) and fluctuations $\delta s^a$ transverse to the trajectory (entropy perturbations) as \cite{GWBM}
\begin{align}
 & \delta \sigma (t,\, \bm{x}) \equiv e_a(t)\, \delta \phi^a(t,\,
 \bm{x})\,, \\
 & s^a(t,\, \bm{x}) \equiv P^a\!_b(t)\, \delta \phi^b(t,\, \bm{x})\,,
\end{align}
where $e_a(t)$ is the unit vector defined by
\begin{align}
 & e^a \equiv \frac{\dot{\phi}^a}{\dot{\sigma}}, \qquad \dot{\sigma}= \sqrt{\dot{\phi}^a \dot{\phi}_a}\,,
\end{align}
and $P_{ab}(t)$ is the projection tensor defined by
\begin{align}
 & P_{ab} \equiv \delta_{ab} - e_a e_b\,.
\end{align}
Note that since $e_a s^a=0$, there are only $A-1$ independent entropy perturbations. 

It follows from the definition of $\delta \sigma$ that  under a change of the time coordinate, $t \to \tilde{t}=t + \delta t$, 
and to linear order in perturbation it transforms as
\begin{align}
 & \delta \sigma \to \delta \tilde{\sigma}= \delta \sigma - \dot{\sigma}
 \delta t \, .
\end{align}
One may construct a corresponding gauge
invariant variable by
\begin{align}
 & \zeta = {\cal R} - \frac{H}{\dot{\sigma}} \delta \sigma \,, \label{Exp:zeta}
\end{align}
where ${\cal R}$ is the curvature perturbation. Then $\zeta$ can be
understood either as the curvature perturbation in the gauge
$\delta \sigma=0$ or as being related to $\delta \sigma$ in the flat gauge, ${\cal R}=0$,
\begin{align}
  \zeta = -\frac{H}{\dot{\sigma}} \delta \sigma_f\,,  \label{Rel:zetadphi}
\end{align}
where 
\begin{align}
 & \delta \sigma_f(\mu,\, {\bm x}) \equiv \delta \sigma (\mu,\, \bm{x})
 \big|_{{\cal R}=0} \,.  \label{flat_gauge}
\end{align}

One may also consider a similar construction with $\delta \sigma$ replaced by 
 $\delta \phi^a$. Then the corresponding gauge invariant variable in the flat gauge, ${\cal R}=0$, is given by
\begin{align}
 & \zeta^a \equiv - \frac{H}{\dot{\phi}^a} \delta \phi_f^a\,.  \label{Def:zetaa}
\end{align}
Using $\zeta^a$ we can express $\zeta$
and $s^a$ as
\begin{align}
 & \zeta = \sum_{a=1}^A \left( \dot{\phi}^a \over \dot{\sigma} \right)^2
 \zeta^a \,, \label{Rel:zetazetaa} \\
 & s^a = - \left( \delta^a\!_b - \frac{\dot{\phi}^a
 \dot{\phi}_b}{\dot{\sigma}^2} \right) \frac{\dot{\phi}^b}{H}
 \zeta^b\,.  \label{Rel:szetaa}
\end{align}
When the background trajectory is predominantly determined by one of
the $A$ scalar fields, $\dot{\phi}^a$, satisfying
$|\dot{\phi}^a| \gg |\dot{\phi}^b|$ for $b \neq a$, the adiabatic curvature
perturbation $\zeta$ approximately agrees with $\zeta^a$.

\subsection{Adiabatic and entropy perturbations from holography}

We will now adapt the action (\ref{QFT_action}) to reflect the split into adiabatic and entropy perturbations.
Using  $\beta^a = \dot{\phi}^a/H$ we obtain,
\begin{align}
 & e^a = \frac{\beta^a}{\beta} , \qquad \beta \equiv \sqrt{\beta^a \beta_a} = 
 \frac{\dot{\sigma}}{H}\,.  \label{Exp:eabeta}
\end{align}
Introducing
\begin{align}
 & {\cal O}_\sigma(\mu,\, \bm{x}) \equiv e^a {\cal O}_a(\mu,\, \bm{x}) \,, \\
 & {\cal O}_{s,a} (\mu,\, \bm{x}) \equiv P_a\!^b {\cal O}_b (\mu,\,
 \bm{x}) \,, \label{Def:Osa}
\end{align}
we may re-express the deformation term as (repeated indices indicate summation)
\begin{align}
  g^a(\mu,\, \bm{x}) {\cal O}_a(\mu,\, \bm{x}) &=  g^a(\mu) {\cal O}_a(\mu,\, \bm{x})
    + \delta \sigma (\mu,\, \bm{x}) {\cal O}_\sigma (\mu,\, \bm{x})
 +  s^a(\mu,\, \bm{x})  {\cal O}_{s,a}(\mu,\, \bm{x})
 \,, \nonumber \\
&= 
 g^a(\mu) {\cal O}_a(\mu,\, \bm{x})
    + \delta \sigma (\mu,\, \bm{x}) {\cal O}_\sigma (\mu,\, \bm{x}) +
    s^{a'}(\mu,\, \bm{x}) \hat{{\cal O}}_{a'} (\mu,\, \bm{x})\,,  \label{Exp:deformation}
\end{align}
where $a'=2, \ldots, A$. Here, we eliminated $s^1$ using $e_a s^a=0$ and 
\begin{align}
 & \hat{{\cal O}}_{a'}(\mu,\, \bm{x})  \equiv {\cal O}_{s,a'}(\mu,\, \bm{x}) - \frac{e_{a'}}{e_1} {\cal
O}_{s,1}(\mu,\,\bm{x})  = \sum_{b=1}^A  \left( \delta_{a'}\!^b -
\frac{e_{a'}}{e_1} \delta_1\!^b \right) {\cal O}_b(\mu,\, \bm{x})\,.
\end{align}

We have thus traded the original operators ${\cal O}_a$ for operators
${\cal O}_\sigma$ and $\hat{\cal O}_{a'}$ that couple directly to the 
adiabatic perturbation $\delta \sigma$ and the $A-1$ entropy perturbations
$s^{a'}$,
\begin{align}
 & S_{\rm QFT}[\chi, g^a; \delta \sigma_f , s^{a'}] \cr
 &= S_{\rm QFT}[\chi, g^a]
 + \int \dd \Omega \left(\delta \sigma_f (\mu,\, \bm{x}) {\cal O}_\sigma (\mu,\, \bm{x})
 +  s^{a'}(\mu,\, \bm{x}) \hat{\cal{O}}_{a'}(\mu,\, \bm{x})\right),  \label{Exp:deformationf}  
\end{align}
where $S_{\rm QFT}[\chi, g^a]$ depends only on the background $g^a(\mu)$.
Since the QFT lives in flat space we replaced  $\delta \sigma$ by
$\delta \sigma_f$ to express the action in terms of gauge invariant
sources (see (\ref{flat_gauge})). 

\subsection{Power spectra}

We are now ready to obtain the relation between the bulk 2-point functions of the adiabatic and entropy perturbations and 2-point functions of the dual QFT. 

Using Eq.~(\ref{Rel:zetadphi}) we may express $\zeta$ derivatives in terms of $ \sigma_f$ ones:
\begin{align}
 &  \frac{\delta W_{\rm
 QFT}}{ \delta \zeta(\bm{x})} = - \beta(\mu) \frac{\delta W_{\rm
 QFT}}{ \delta \sigma_f(\bm{x})} \,.  \label{WTcp2}
\end{align} 
Taking further derivatives and using Eq.~(\ref{2pt})  and the results in the previous subsection we get
\begin{align}
 W^{(2)}_{11} (\bm{x}_1,\, \bm{x}_2) &  =\frac{\delta^2 W_{\rm
 QFT}}{ \delta \zeta(\bm{x}_1) \delta \zeta(\bm{x}_2)} =  
 \beta^2(\mu) \frac{\delta^2 W_{\rm QFT}}{ \delta \sigma_f(\bm{x}_1)
 \delta \sigma_f(\bm{x}_2)} \nonumber \\
 &=2  \beta^a(\mu) \beta^b(\mu)\, {\rm Re} \left[ \langle {\cal
 O}_a(\bm{x}_1) {\cal O}_b(\bm{x}_2) \rangle_\mu \right]\,, \label{Exp:W211} \\
 W^{(2)}_{1a'} (\bm{x}_1,\, \bm{x}_2) & =  \frac{\delta^2 W_{\rm
 QFT}}{ \delta \zeta(\bm{x}_1) \delta s^{a'}(\bm{x}_2)} =  - 
 \beta(\mu)  \frac{\delta^2 W_{\rm
 QFT}}{ \delta \sigma_f(\bm{x}_1) \delta s^{a'}(\bm{x}_2)} \nonumber \\
 & =-2 \beta^a(\mu) \left( \delta_{a'}\!^b -
\frac{\beta_{a'}}{\beta_1} \delta_1\!^b \right) \, {\rm Re} \left[ \langle {\cal
 O}_a(\bm{x}_1) {\cal O}_b(\bm{x}_2) \rangle_\mu \right]\,, \label{Exp:W21ad} \\
W^{(2)}_{a'b'} (\bm{x}_1,\, \bm{x}_2) &=\frac{\delta^2 W_{\rm
 QFT}}{ \delta s^{a'}(\bm{x}_1) \delta s^{ab}(\bm{x}_2)} \nonumber \\ 
&=2  \left( \delta_{a'}\!^a -
\frac{\beta_{a'}}{\beta_1} \delta_1\!^a \right)  \left( \delta_{b'}\!^b -
\frac{\beta_{b'}}{\beta_1} \delta_1\!^b \right)  {\rm Re} \left[ \langle {\cal
 O}_a(\bm{x}_1) {\cal O}_b(\bm{x}_2) \rangle_\mu \right]\,. \label{Exp:W2adbd}
\end{align}
Thus the two-point function of ${\cal O}_{\Phi^a}$, which is dual to $\Phi^a$,
is now expressed by the two-point function of ${\cal O}_a$. An additional
minus sign is introduced by taking into account the relative minus sign
between the two-point functions in QFT$_+$ and QFT$_-$.

The (equal time) 2-point function of $\zeta$ and $s^{a'}$ can be obtained from 
 the inverse matrix of $W^{(2)}_{ab}(\bm{x}_1,\, \bm{x}_2)$, see Eq.~(\ref{Exp:PP}).  In Fourier space and restricted to $d=3$ from now on
\begin{align}
 & \langle \Phi_a(\bm{k}_1) \Phi_b(\bm{k}_2) \rangle_{\rm conn}  =
 (2\pi)^3 \frac{2 \pi^2}{k_1^3} {\cal P}_{ab}(k_1)  \delta (\bm{k}_1 + \bm{k}_2)
\end{align}
where 
\begin{align}
 & \hat{W}^{(2)\,-1}_{ac}(k) \hat{W}^{(2)}_{cb}(k) = \delta_{ab}\,,
\end{align}
and 
\begin{align}
  & W^{(2)}_{ab}(\bm{x}_1,\, \bm{x}_2) = \int \frac{\dd^3 \bm{k}}{(2\pi)^3}\, e^{i
 \sbm{k} \cdot (\sbm{x}_1- \sbm{x}_2)}\, \hat{W}_{ab}^{(2)} (k)\,.
\end{align}
The scalar power spectra are now given by
\begin{align}
 & {\cal P}_{ab}(k) = \frac{k^3}{2 \pi^2} \hat{W}^{(2)\,-1}_{ab}(k)\,.
\end{align}

Once the beta function $\beta^a$ and the correlators of ${\cal O}_a$ are
derived by solving the RG flow, these formulae may be use to
compute the power spectra for the adiabatic and entropy perturbations.

\section{Power spectra for the separable example}  \label{Sec:example}

In this section we explicitly compute the power spectra for the separable example.

\subsection{Formulae for the separable case}

The beta functions $\beta^a(\mu)$ may be readily computed from Eq.~(\ref{Sol:gs}),
\begin{align}
 & \beta^a(\mu) = \frac{4 \left( \mu/ \mu_a \right)^{\lambda_a}}{\left[  1+  \left( \mu/\mu_a
 \right)^{\lambda_a} \right]^2} 
 \beta^a(\mu_a)\,,   \qquad \beta^a(\mu_a) 
 =-\frac{\lambda_a^2}{2 \tilde{C}_a}\,.
 \label{Exp:betasms}
\end{align}

The two-point function of the operators ${\cal O}_a$ may be computed using the same steps 
as in the single field case~\cite{BMS,McFadden:2013ria,JYcsv}  leading to
\begin{align}
 \langle  {\cal O}_a(\bm{x}_1) {\cal O}_b(\bm{x}_2) \rangle_\mu
  &= \delta_{ab}  \left[ 1+ \left( \mu \over \mu_a \right)^{-\lambda_a} \right]^4 
  \left[ 1 + (\mu_a r)^{\lambda_a} \right]^{-4} \frac{c_a
 \mu^{2\lambda_a}}{r^{2(3-\lambda_a)}} \cr
  & = \delta_{ab} \left[ \beta^a(\mu_a) \over \beta^a(\mu) \right]^2 {\cal
 F}_a(r) \,, \label{Exp:O2p}
\end{align}
where we introduced
\begin{align}
 & {\cal F}_a(r) \equiv \frac{ 16 c_a \mu_a^{2\lambda_a}}{r^{2(3-\lambda_a)}[ 1 +  (\mu_a
 r)^{\lambda_a}]^4}\,.
\end{align}
Here, $r$ denotes $r \equiv |\bm{x}_1 - \bm{x}_2|$. In obtaining this
result we fixed a number of integration constants by requiring the
correlators behave appropriately in the limit $\lambda \to 0$, see the
discussion in Sec.~2.2 of Ref.~\cite{BMS}.

\subsection{Power spectra for the adiabatic and entropy perturbations}

Inserting Eqs.~(\ref{Exp:betasms}) and (\ref{Exp:O2p}) into
Eqs.~(\ref{Exp:W211})-(\ref{Exp:W2adbd}) we may now compute
$W^{(2)}_{ab}(x_1,\, x_2)$. Then, computing the inverse matrix $\hat{W}^{(2)\,-1}_{ab}(k)$
leads to the power spectra for the adiabatic and entropy
perturbations for the separable RG flow. We relegate this computation to appendix \ref{comp}.
Since the RG equations are separable the power spectra for the adiabatic and entropy perturbations may be derived more easily
by using the power spectra of $\zeta^a$.

Repeating a similar computation to the one for the correlator of
$\zeta$, we obtain the correlator of $\zeta^a(\bm{x})$ as
\begin{align}
 & \langle \zeta^a(\bm{x}_1) \zeta^b(\bm{x}_2) \rangle = W^{(2)\,-1}_{\zeta^a \zeta^b}
 (\bm{x}_1,\, \bm{x}_2)\,,
\end{align}
with
\begin{align}
 &  W^{(2)}_{\zeta^a \zeta^b} (\bm{x}_1,\, \bm{x}_2) = 2 {\rm Re} \left[\beta^a(\mu) \beta^b(\mu) \langle {\cal O}_a(\bm{x}_1) {\cal
 O}_b(\bm{x}_2) \rangle_\mu  \right]\,. \label{w2}
\end{align}

Since the correlators of ${\cal O}_a$ are diagonal (see (\ref{Exp:O2p})), one may readily compute the inverse matrix
$W^{(2)\,-1}_{\zeta^a \zeta^b} (\bm{x}_1,\, \bm{x}_2)$ to obtain the
power spectrum of $\zeta^a$,\footnote{To compute the Fourier transform of (\ref{w2}) we expand the
$r$-dependent part of ${\cal F}_a(r)$ as
$$
r^{-2(3-\lambda_a)}[ 1 +  (\mu_a r)^{\lambda_a}]^{-4}  =
  \sum_{m=0}^\infty \frac{(m+3)!}{3!\,m!} \left( - (\mu_a)^{\lambda_a}
  \right)^m r^{-6+\lambda_a(m+2)} \,,
$$
and use
$$
\int \dd^3 \bm{x} r^{-6+(m+2) \lambda_a} e^{-i \sbm{k} \cdot \sbm{x}} =
\frac{\pi^2}{12} k^{3-(m+2) \lambda_a} (1 + {\cal O}(\lambda_a)).
$$}
\begin{align}
 {\cal P}_{\zeta^a \zeta^a} (k)
 &\simeq \frac{3}{16 \pi^4} \frac{1}{
\left(\beta^a(\mu_a)\right)^2 c_a}  \left( k \over \mu_a
\right)^{- 2 \lambda_a} 
 \left[ 1 +  \left( k \over \mu_a
 \right)^{ \lambda_a}  \right]^4 \!\! \left(1 + {\cal O}(\lambda_a) \right)
 \,. \label{Exp:Pzetaa}
\end{align}
In the single field case this is the scalar power spectrum. Note that
the $\mu$ dependence cancels out  between the explicit $\beta^a(\mu)$
factors and the $1/(\beta^a(\mu))^2$ factor in (\ref{Exp:O2p}). Thus the
power spectrum in the single field case is indeed conserved. One may
check that this agrees exactly with corresponding result in cosmological
perturbation theory provided one uses the standard AdS/CFT normalisation
for  the operator (i.e. set $c_a=12 /(\pi^2 \kappa^2$)) \cite{BMS}.

We may now compute the auto-correlation of  $\zeta$ using Eq.~(\ref{Rel:zetazetaa}): 
\begin{align}
 & {\cal P}_{11} (k) = \sum_{a=1}^A \left( \beta^a(\mu) \over \beta(\mu)
 \right)^4 {\cal P}_{\zeta^a \zeta^a}(k)\,.  \label{Exp:Pzeta}
\end{align}
Similarly, using Eq.~(\ref{Rel:szetaa}), we may obtain the
cross-correlation between $\zeta$ and $s^{a'}$ and the auto-correlation
of $s^{a'}$. For instance, the auto-correlation for $s^{a'}$ is given
by
\begin{align}
 & {\cal P}_{a'a'}(k) = \left(\beta^{a'}(\mu) \right)^2 \! \left[  \left\{ 1 -
 \left( \beta^{a'}(\mu) \over \beta(\mu) \right)^2 \right\}^2 \!\!
 {\cal P}_{\zeta^{a'} \zeta^{a'}} (k) +
 \sum_{c \neq a'} \left( \beta^c(\mu) \over \beta(\mu) \right)^4
 {\cal P}_{\zeta^c \zeta^c}(k)  \right]. \label{Exp:Psad}
\end{align}
These results agree with the ones obtained by directly inverting $W^{(2)}_{ab}(k)$, see Appendix \ref{comp}, as it should be.

We now come to our main point: the power spectrum (\ref{Exp:Pzeta}) depends on $\mu$ and thus depends on
time in cosmology.  One should contrast this with the  case of a deformation by a single
operator where (as we mentioned above) the power spectrum is $\mu$ independent~\cite{JYcsv}.
This is consistent with the results obtained by
standard cosmological perturbation theory which show a time variation of $\zeta$ at large
scales in the presence of the entropy perturbation~\cite{GWBM}.  The $\delta N$ formalism~\cite{Salopek:1990jq, Starobinsky:1986fxa, Sasaki:1995aw, Sasaki:1998ug, Lyth:2004gb} provides a powerful way to compute the
primordial curvature perturbation $\zeta$ for multi-field models. Using this formalism, the curvature
perturbation $\zeta$ has been computed in Ref.~\cite{JYF} for the cosmology 
dual to the RG flow (\ref{Eq:betam}) and the power spectrum of $\zeta$ has indeed been found to agree exactly 
with Eq.~(\ref{Exp:Pzeta}).

When $\lambda_1$ is smaller than $\lambda_{a'}$ with $a' \neq 1$, $|\beta^1(\mu)|$ becomes much bigger than
$|\beta^{a'}(\mu)|$ in the UV. Then, the curvature perturbation $\zeta$ agrees with
$\zeta^1$ and the power spectrum ${\cal P}_{11}(k)$ is approximately given by the constant
value,
$$
{\cal P}_{\zeta \zeta}(k) \simeq {\cal P}_{\zeta^1 \zeta^1}(k).
$$
By contrast, the power spectrum ${\cal P}_{a' a'}(k)$ still
depends on $\mu$ in this limit and vanishes as
$$
 {\cal P}_{a' a'}(k) \simeq  \left(\beta^{a'}(\mu) \right)^2 \left[
 {\cal P}_{\zeta^1 \zeta^1}(k) + {\cal P}_{\zeta^{a'} \zeta^{a'}}(k) \right], \qquad \mu \to \infty \,.
$$
Notice that in the case all $\beta^a(\mu)$ are proportional to each other,
the power spectrum of $\zeta$ is conserved. This is because the
evolution of $\phi^a$ can be described by a single degree of freedom
after an appropriate rotation in the field space.

As in the singe field case~\cite{BMS}, the power spectrum is
red-tilted, where the spectral index for $\zeta^a$ is given by $n_{s,a} -1 = - 2 \lambda_a$ in the 
long wavelength limit, $k \ll \mu_a$,  while it is blue-tilted, $n_{s,a}-1=2 \lambda_a$
in the opposite limit, $k \gg  \mu_a$. This is due to the fact that 
the potential given in Eq.~(\ref{Exp:Vphi}) has negative curvature near
the IR fixed point (which corresponds to the early times in cosmology)  
while it  has positive
curvature near the UV fixed point (which corresponds to the late times in cosmology) .

\section{Concluding remarks} \label{Sec:conclusions}
In this paper we initiated the study of multi-field inflation using holography.
We studied a class of models where the Universe evolves from one de Sitter phase to another.
Cosmological evolution maps  to inverse RG flow in holography and  thus the dual quantum field 
theory for this class of models has two fixed points. Multi-field inflation corresponds to an RG flow driven by a
multitude of nearly marginal operators. In this paper we analysed a special case: the deformation operators do not couple to each 
at leading order, i.e. there are no non-zero three-point functions that involve more than one type of operator.
Under this condition one can compute the relevant QFT correlators
exactly in the same way as in the corresponding
single field models. 

We explicitly computed holographically the power spectra of the adiabatic curvature and
entropy perturbations for the case of several deformation operators and recovered one of the hallmarks of 
multi-field inflation: the curvature perturbation $\zeta$ is not conserved in the presence of entropy perturbations.
The curvature perturbation is conserved when the entropy perturbations
are set to zero or when the model can be described by a single field
after a rotation in field space. Furthermore, in the UV (=late time) limit  the power spectrum approaches a constant
value and it is solely determined by the operator which is closest to be marginal
(i.e. has the minimum value of $\lambda_a$).

In this paper, we computed only the power spectra. It would be interesting to extend the analysis to 
higher point functions. However,  as pointed out in Ref.~\cite{JYcsv},  already the single field case is 
subtle.  The curvature perturbation 
$\zeta$ in such models should be conserved at large scales as a consequence of general 
covariance but the computation of the bispectrum in \cite{JYcsv} yielded a result that depended on $\mu$.
One would need to first resolve this issue before moving to higher point function in  the multi-field case.

Here, we focused on the simplest (but very special) QFT model with many
nearly marginal operators, namely the case where there is no operator mixing
along the RG flow. It would be very interesting to extend the analysis to the generic case. It would also be interesting to 
present holographic versions of multi-field models that are already discussed in the cosmology literature.
We leave such studies for future work.

\acknowledgments
We would like to thank Paul McFadden for discussions. Y.~U. would like to
thank University of Southampton, University of Padua and CERN for their
hospitalities during the period of this work. J.~G. and Y.~U. are partially supported by MEC FPA2010-20807-C02-02, AGAUR
2009-SGR-168, MEC FPA2013-46570-C2-2-P, AGAUR 2014-SGR-1474, and CPAN CSD2007-00042 Consolider-Ingenio 2010. 
K.S. acknowledges support from a grant of the John Templeton
Foundation. The opinions expressed in this publication are those of the authors and do not
necessarily reflect the views of the John Templeton Foundation. 
K.S. would also like to thank the Aspen Center for Physics for the
hospitality during the final stages of this work and acknowledge partial
support by National Science Foundation Grant No. PHYS-1066293. Y.~U. is
supported by ECOST-STSM-MP1210-241113-037164 and JSPS No.~21244033.

\appendix

\section{Analytic continuation} \label{anal}

We review in this appendix the analytic continuations needed using an
example the case of a massless scalar field in four (bulk)
dimensions. This example was discussed in detail in
Ref.~\cite{Maldacena02} so we will simply borrow the results from
there. This appendix also serves to illustrate that the analytic
continuations in Ref.~\cite{Maldacena02} and Ref.~\cite{MS_HC09},
despite apparent differences, are equivalent.

The renormalised on-shell action for a massless scalar in EAdS is given
by
\begin{align}
 S_{AdS\ {\rm ren}} &= - \frac{1}{\kappa^2} \int \frac{\dd^3 \bm{k}}{2 \pi^3}
 \frac{1}{2} R_{AdS}^2 k^3 \phi_0(-k) \phi_0(k) \cr
 & = - \int \dd^3 \bm{k} \langle
 O(k) O(-k) \rangle \phi_0(-k) \phi_0(k)\,, \label{AdS} 
\end{align}
where 
$\phi_0$ is value of $\phi$ at the conformal boundary of AdS and the source for the operator $O$.
This is obtained by directly evaluating the on-shell action and using holographic renormalisation \cite{Skenderis:2002wp}  (see Eq.~(5.4) of Ref.~\cite{Maldacena02}). Here (as in the main text) we adopt the ``supergravity normalisation''
where there is an overall factor of $1/\kappa^2$ in front of the action.  The 2-point function of dual operator is then given by
\begin{equation} \label{M_norm}
\langle O(k) O(-k) \rangle = - \frac{\delta^2 S_{AdS\ {\rm ren}}}{\delta
 \phi_{0} (k) \delta \phi_{0} (-k)}  =  \frac{1}{2}
 \frac{R_{AdS}^2}{\kappa^2} k^3  \,.
\end{equation}

A direct evaluation of the on-shell action in de Sitter (again normalised with an overall $1/\kappa^2$) yields \cite{Maldacena02}
\begin{equation} \label{dS}
{\rm Re} (i S_{dS}) = \frac{1}{\kappa^2} \int \frac{\dd^3 \bm{k}}{2 \pi^3}  \frac{1}{2} R_{dS}^2 k^3 \phi_0(-k) \phi_0(k) \,.
\end{equation}
This is related to (\ref{AdS}) by 
\begin{equation} \label{R_cont}
R_{AdS} = -i R_{dS} \,.
\end{equation}
Then one can compute the bulk 2-point function using the wave function $\psi_{\rm bulk}[\phi_{0}]=\exp (i S_{dS})$ to obtain
\begin{equation} \label{FF_M}
\langle \phi(k) \phi(-k) \rangle = -\frac{1}{2 {\rm Re} \langle O(k) O(-k) \rangle}\Big|_{R_{AdS} = -i R_{dS}} = \frac{\kappa^2}{ R_{dS}^2} \frac{1}{k^3}\,. 
\end{equation}

In Ref.~\cite{MS_HC09} the bulk in-in correlators were computed using
the in-in formalism and compared with the corresponding correlators in
the dual QFT. In this section, the AdS and dS radii were set to one and the following holographic formula was obtained
\begin{equation} \label{McS_norm}
\langle \phi(p) \phi(-p) \rangle = -\frac{1}{2 {\rm Im} \langle O(-i p) O(i p) \rangle}\Big|_{\kappa^2 \to -\kappa^2} = \kappa^2 \frac{1}{{\rm Im} (-i p)^3} = \kappa^2 \frac{1}{p^3}\,, 
\end{equation}
which indeed agrees with Eq.~(\ref{FF_M}) upon setting $R_{dS}=1$ there. 

The apparent difference between the two prescriptions is due to fact that the three dimensional correlator in Eq.~(\ref{M_norm}) is on 
$\mathbb{R}^3$  with metric 
$\dd s^2 = R_{AdS}^2 \dd x_i \dd x_i$, while that in
Eq.~(\ref{McS_norm}) is on $\dd s^2 = \dd x_i \dd x_i$. This implies that $p= R_{AdS} k$ and under the continuation in Eq.~(\ref{R_cont}), $p \to -i p$. This then converts the real part in Eq.~(\ref{FF_M}) to the imaginary part in Eq.~(\ref{McS_norm}).
Once we set $R_{dS}=1$ the overall minus in the correlator due to the overall factor of $R_{dS}^2$ in Eq.~(\ref{M_norm}) is now
accounted for by taking $\kappa^2 \to -\kappa^2$.  

In this example, the entire effect of the analytic continuation is to produce an overall minus sign. In more general cases the action of the analytic continuation is more non-trivial and produces additional signs, see Ref.~\cite{McFadden:2013ria} for an example. If one uses the conventions of Ref.~\cite{Maldacena02} one must keep explicitly the factors of $R_{dS}$ in order to do correctly the analytic continuation.
In the cases we discuss in this paper where we only compute the leading order terms (in the deformation parameters) the effect of the analytic continuation is to only produce an overall 
sign. 

Finally, we note that the analytically continued correlators that enter the holographic formulae of the  power spectrum also have a meaning in the original QFT$_-$ without any analytic continuation: they are related with the spectral density associated with the 2-point function of the energy momentum tensor~\cite{McFadden:2013ria}.

\section{Inverting  $W^{(2)}_{ab}$} \label{comp}
In this appendix, we derive the power spectra for the adiabatic and
entropy perturbations by computing the inverse matrix
$\hat{W}^{(2)\,-1}_{ab} (k)$. For simplicity, we consider the two field
case where the structure constant $C_{abc}$ satisfies
Eq.~(\ref{Cond:diagonal}) and hence the RG equation becomes separable.
Inserting Eqs.~(\ref{Exp:betasms}) and (\ref{Exp:O2p}) into
Eqs.~(\ref{Exp:W211})-(\ref{Exp:W2adbd}), we obtain
\begin{align}
 & \hat{W}^{(2)}_{11} (k) =  2 
 \sum_{a=1}^2 \left( \beta^a(\mu_a)\right)^2 \hat{{\cal F}}_a(k) \,, \label{Exp:W211kap}\\
 & \hat{W}^{(2)}_{12} (k) =   2 \beta_{2}(\mu) \left[
 \left(\beta^1(\mu_1) \over \beta^1(\mu) \right)^2 \hat{{\cal F}}_1(k) - \left(
 \beta^2(\mu_2) \over \beta^2(\mu) \right)^2 \hat{\cal F}_2(k) \right]\,, \label{Exp:W21adkap} \\
 & \hat{W}^{(2)}_{22}(k) = 2 \left[ \left( \beta^2(\mu) \over
 \beta^1(\mu) \right)^2 \left(\beta^1(\mu_1) \over
 \beta^1(\mu) \right)^2 \hat{\cal F}_1(k) +  \left(\beta^2(\mu_2) \over
 \beta^2(\mu) \right)^2 \hat{\cal F}_2(k) \right]\,, \label{Exp:W2adbdkap}
\end{align}
where $\hat{\cal F}_a(k)$ denotes the Fourier mode of ${\cal F}_a(r)$, given by
\begin{align}
 & \hat{\cal F}_a(k) = \frac{4}{3} \pi^2 c_a k^3 \left( k \over \mu_a
 \right)^{2 \lambda_a} \left[ 1 +  \left( k \over \mu_a
 \right)^{\lambda_a}  \right]^{-4} \left(1 + {\cal O}(\lambda_a) \right)\,. \label{Exp:hatfk}
\end{align}
Using $\hat{W}^{(2)}_{ab}(k)$, the  $2 \times 2$ matrix
$\hat{W}^{(2)\,-1}_{ab}(k)$ is given by
\begin{align}
 & \hat{W}^{(2)\,-1}_{11}(k) = \frac{\hat{W}^{(2)}_{22}(k)}{\det
 \hat{W}^{(2)}(k)}\,, \label{Exp:W2I11f} \\
 & \hat{W}^{(2)\,-1}_{22}(k) = \frac{\hat{W}^{(2)}_{11}(k)}{\det \hat{W}^{(2)}(k)}\,, \\
 & \hat{W}^{(2)\,-1}_{12}(k) = \hat{W}^{(2)\,-1}_{21}(k)
  = - \frac{\hat{W}^{(2)}_{12}(k)}{\det \hat{W}^{(2)}(k)}\,, \label{Exp:W2I22f}
\end{align}
with
\begin{align}
 & \det \hat{W}^{(2)}(k) \equiv \hat{W}^{(2)}_{11}(k)
 \hat{W}^{(2)}_{22}(k) - \left(\hat{W}^{(2)}_{12}(k) \right)^2\,, \label{Def:W}
\end{align}
where we used $\hat{W}^{(2)}_{12}(k)=\hat{W}^{(2)}_{21}(k)$.

Inserting Eqs.~(\ref{Exp:W211kap})-(\ref{Exp:W2adbdkap}) into
Eq.~(\ref{Def:W}) we obtain
\begin{align}
 & \det \hat{W}^{(2)}(k) = 4 \hat{{\cal F}}_1(k)
 \hat{\cal F}_2(k)  \left( \beta^1(\mu_1) \beta^2(\mu_2) \over \beta^2(\mu)
 \right)^2 \left(\beta(\mu) \over \beta^1(\mu) \right)^4 \,. \label{Exp:W}
\end{align}
Using Eqs.~(\ref{Exp:W211kap}), (\ref{Exp:W2adbdkap}), and
(\ref{Exp:W}), we obtain the auto-correlations of $\zeta$ and $s^2$ as
\begin{align}
 & {\cal P}_{11}(k) =  \frac{k^3}{4 \pi^2} \sum_{a=1}^2 \left( \frac{\beta^a(\mu)}{\beta(\mu)} \right)^4 \frac{1}{\left(
 \beta^a(\mu_a) \right)^2 \hat{{\cal F}}^a(k)}\,, \\
 & {\cal P}_{22}(k) =  \frac{k^3}{4 \pi^2} \frac{\left( \beta^1(\mu) \right)^4 \left(\beta^2(\mu)
 \right)^2}{\left(\beta(\mu)\right)^4} \sum_{a=1}^{2} \frac{1}{\left(
 \beta^a(\mu_a) \right)^2 \hat{{\cal F}}^a(k)}\,,
\end{align}
which agree with Eqs.~(\ref{Exp:Pzeta}) and (\ref{Exp:Psad}),
respectively.


\end{document}